\documentclass[article,twocolumn,aps,superscriptaddress,longbibliography]{revtex4-1}

\newcommand{\vct}[1]{{\bf #1}}
\newcommand{\eq}[1]{Eq.~(\ref{eq:#1})}
\newcommand{\fig}[1]{Fig.~\ref{fig:#1}}

\usepackage{graphicx}
\usepackage{dcolumn}
\usepackage{bm}
\usepackage[usenames,dvipsnames]{color}
\usepackage{xcolor}
\usepackage{mathrsfs}
\usepackage{scalefnt}

\definecolor{cgreen}{RGB}{0,158,115}
\definecolor{cblue}{RGB}{86,180,233}
\definecolor{cred}{RGB}{229,30,16}
\definecolor{cpurple}{RGB}{148,0,211}
\definecolor{cyellow}{RGB}{240,228,66}
\definecolor{corange}{RGB}{230,159,0}
\definecolor{munsell}{rgb}{0.94, 0.8, 0.0}
\let\xxxhat\hat

\renewcommand{\hat}[1]{{\boldsymbol {\xxxhat {#1}} }}
\renewcommand{\vec}[1]{\boldsymbol {#1}}

\usepackage{graphicx}
\usepackage{dcolumn}
\usepackage{bm}
\usepackage{epstopdf}
\usepackage{float}
\usepackage[colorlinks=true, linkcolor=blue, citecolor=cyan, urlcolor=purple]{hyperref}
\usepackage{amsmath}
\usepackage{xcolor}
\usepackage{comment}
\usepackage[normalem]{ulem}

\begin{document}
\title{Fermi Polaron in Atom-Ion Hybrid  Systems }

\author{Renato Pessoa}
\affiliation{Instituto de F\'{\i}sica,
Universidade Federal de Goi\'as - UFG,
74001-970 Goi\^ania, GO, Brazil} 

\author{S.~A.~Vitiello}
\affiliation{Instituto de F\'{\i}sica Gleb Wataghin,
Universidade Estadual de Campinas - UNICAMP,
13083-970 Campinas, SP, Brazil}

\author{L.~A.~Pe\~na Ardila}
\email{luisaldemar.penaardila@units.it}
\affiliation{Dipartimento di Fisica, Università di Trieste, Strada Costiera 11, I-34151 Trieste, Italy}
\affiliation{School of Science and Technology, Physics Division, University of Camerino, Via Madonna delle Carceri, 9B - 62032 (MC) , Italy}

\date{\today}

\begin{abstract}

Atom-ion hybrid systems are promising platforms for the quantum simulation of polaron physics in certain quantum materials. Here, we investigate the ionic Fermi polaron, a charged impurity in a polarized Fermi bath, at zero temperature using quantum Monte Carlo techniques. We compute the energy spectrum, residue, effective mass, and structural properties. Significant deviations from field-theory prediction occur in the strong coupling regime due to large density inhomogeneities around the ion. We observe a smooth polaron-molecule transition in contrast with the neutral case. This study provides insights into solid-state systems like Fermi exciton polarons in thin semiconductors and quantum technologies based on atom-ion platforms.
\end{abstract}

\maketitle

 \textit{Introduction --} Impurities interacting with a quantum many-body environment lead to the formation of quasiparticles termed polarons as the mobile impurity entangles with the virtual quantum excitations in the medium~\cite{bay91,MahanBook,lan48}. 
 Polarons were first introduced in the context of electrons embedded in polar crystals. The high level of controllability in ultracold degenerate quantum mixtures has theoretically and experimentally inspired the study of a quantum analog of the solid-state polarons. These quasiparticles arise from the interaction between impurities and the low-energy excitations of the quantum gas. Experimental realization of both Bose~\cite{Jorgensen2016,Hu2016,Ardila2019,Yan2020,Skou2021} and Fermi polarons~\cite{Massignan2011,Schmidt_2018,Scazza2022,lou18,Ness2020} were achieved using alkaline atomic species employing spectroscopy and interferometric protocols to characterize these quasiparticles in cold atoms setups~\cite{Vale&Zwierlein2021}.
 
Cold ion-atom hybrid systems have emerged as a robust field at the crossroads of two well-consolidated fields, ultracold quantum gases and ion-trapped systems, offering potential and promising applications in quantum technologies, particularly quantum sensing and quantum computation,  as well as in quantum simulation~\cite{Tomza2019Review}. In this context, combining quantum impurities and atom-ion hybrid systems may provide a powerful platform for analog quantum simulation~\cite{JachymskiPRR2020,Bissbort2013,Dehkharghani2017} and computation~\cite{Kokail2019,rmp-ions,Blatt2012,Schneider_2012,Daley2022}. For example,  impurities can be used as a probe for thermometry~\cite{Mitchison2020,Oghittu2022},  for implementing hybrid quantum information platforms that rely on the individual control and manipulation of trapped ions ~\cite{Doerk2010,Secker2016},  and for the optimization of ion logic gates for quantum computation purposes that may be jeopardized due to heating as the number of gates increases, in which a quantum host gas may serve as a coolant to mitigate possible heating effects
~\cite{Kielpinski2002,Egan2021,Postler2022}. Furthermore, the study of ultracold hybrid atom-ion systems paves the way to exploring nonradiative scattering processes \cite{aym11,hal13,xin22,moh21,dut18,nir24,jos22,pan20cs,pan20,cot00,zha09,zha11,pur19,li19pra,hud18,pur18,pur17,din22,aha17,cot02,mak03,cot16,gac16,cot18,pin23,ben21}, which has led to the development of an advanced algorithm for quantum scattering of atoms \cite{xin23}.

Because of the hierarchy in the natural energy and length scales between solid-state crystal and ultracold atomic systems, impurities in hybrid atom-ion systems serve as a test bed for currently unattainable regimes in solid-state~\cite{Bissbort2013}. Despite the extensive theoretical and experimental work on polarons in ultracold quantum gases, the degeneracy realm for the atom-ion mixture is still unreachable within current technology~\cite{lou22}  but is rapidly advancing.  The charged impurity case offers advantages in transport studies with respect to its neutral counterpart, as demonstrated by employing external electric fields to investigate ionic impurity transport~\cite{die21}. These techniques hold promise for exploring polaron properties, including effective mass and nontrivial diffusion properties, in the quantum degenerate regime. In addition, spatial and temporal correlation in polarons can be traced with high precision using pulsed ionic microscopes~\cite{vei21}, enabling access to the complicated out-of-equilibrium polaron dynamics~\cite{Skou2021,Drescher19,ArdilaPRA2021}. The recent experimental capability to manipulate the atom-ion scattering length through Feshbach resonances \cite{wec21} has triggered extensive theoretical investigations of impurities~\cite{cot02,Massignan2005,Schurer2017}, and polarons in hybrid atom-ion systems, in both Bose-Einstein condensates~\cite{Casteels2011,ast21,chr21,Astrakharchik2023} and Fermi gases~\cite{chr22,mysliwy2023longrange}. The key feature of this new polaronic flavor is the lack of length scale separation, given that the typical coherence length is comparable to the potential range. This implies that chemistry associated with few-body bound states may play a crucial role in the formation of many-body bound states or mesoscopic polarons~\cite{Perez2021}. As a result, in certain regimes, especially in the strongly correlated regime, the interplay between few -- and many-body physics is responsible for density deformations, which are substantially underestimated by traditional techniques. Thus, long-ranged polarons featuring large backaction effects of the quantum medium in the presence of a strongly coupled impurity require \textit{ab initio} techniques.

Using \textit{ab initio} quantum Monte Carlo (QMC) techniques within the fixed node approximation, we characterize the polaron by computing its quasiparticle properties and the enhanced density extracted from the atom-ion pair correlation functions. For strong atom-ion interaction, the polaron properties change
dramatically due to the density enhancement around the ion. Furthermore, the polaron-molecule transition exhibited in the neutral case also appears in our spectrum; however, the transition is
continuous, which recently was measured for the neutral
case~\cite{Ness2020} only for finite temperature and finite impurity
density. Our method includes the nonperturbative deformation of the bath and therefore, incorporates the nontrivial backaction effects needed for characterizing the polaron in the presence of strong interactions arising from the complex atom-ion potential.  \\

\begin{figure}[htp]
\centering
\includegraphics[width=8.5cm]{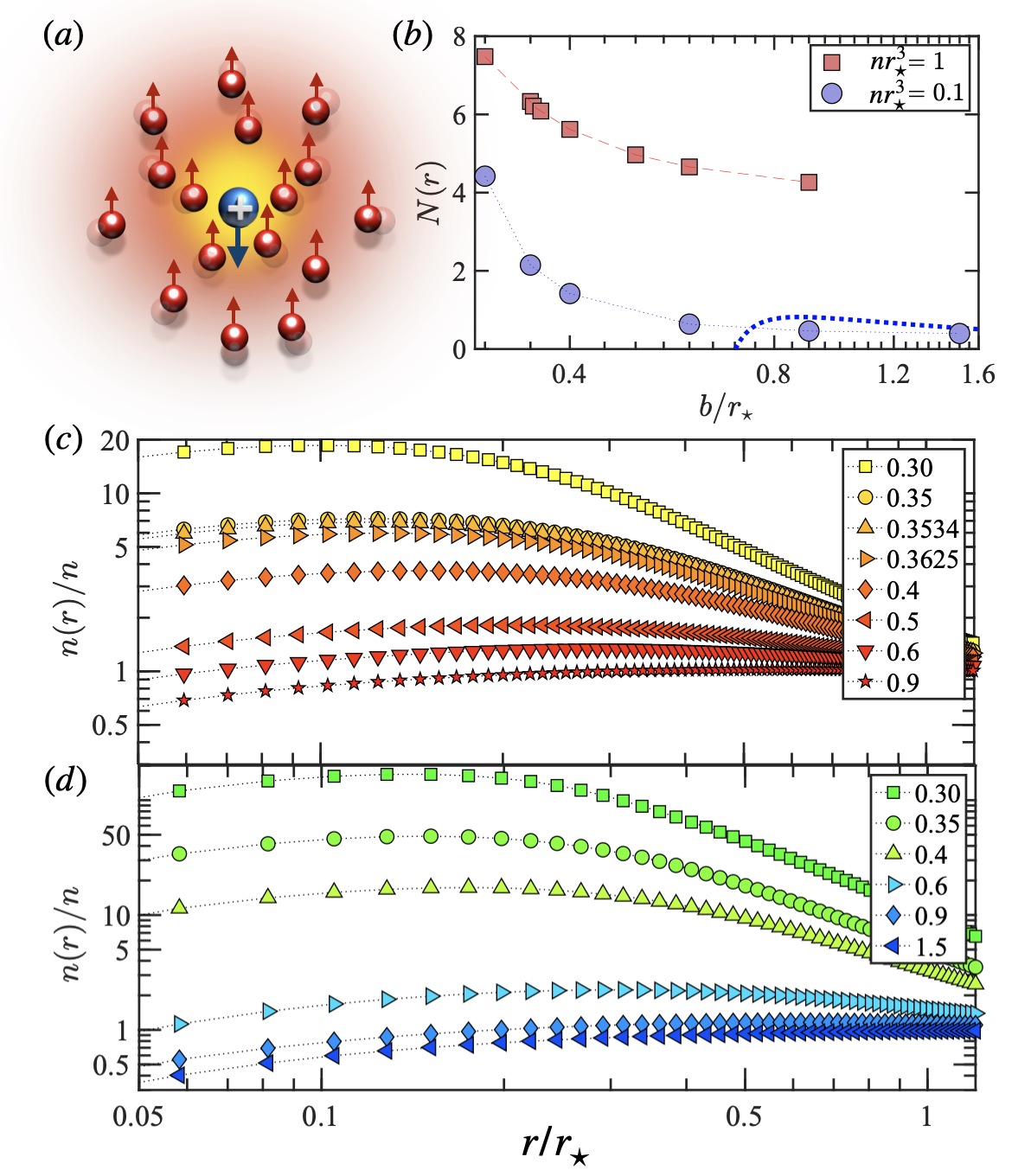}
\caption{(a) Ionic impurity $\mathrm{I}$ with spin $\downarrow$ immersed in a polarized Fermi gas $\uparrow$.  (b) Average number of fermions in the neighborhood of the ion enclosed in a sphere of radius $r_{\star}$ for two unperturbed densities,: $nr_{\star}^3=1$ (red squares) and $nr_{\star}^3=0.1$ (blue circles). The dotted blue line indicates the theoretical prediction for both the neutral Fermi polaron valid in the low-density regime and weak coupling, $b\gg r_{\star}$~\cite{Scazza2022}. Density profile $n(r)$ as a function of distance for the given values of $b$ indicated by different symbols at densities (c) $nr_{\star}^3=1$  and (d) $nr_{\star}^3=0.1$, respectively. For large distances $r \gg r_{\star}$, the density converges to the unperturbed density, $n(r)/n \rightarrow 1$.
}
\label{fig:fig1}
\end{figure}
\textit{Model and atom-ion interaction --} We investigate  a single mobile ionic impurity with mass \(m_\mathrm{I}\) and down spin, interacting with an oppositely spin-polarized, noninteracting Fermi gas of mass \(m\) at zero temperature, as illustrated in Fig.~\ref{fig:fig1}(a). The system is characterized by an \textit{unperturbed} density, defined as Fermi gas density in the absence of the ion, $n=N/V=\left(2m\epsilon_{\mathrm{F}}/\hbar^2\right)^{3/2}/6\pi^2$,  where $\epsilon_{\mathrm{F}}$ is the Fermi energy, $N$ and $V$ are the number of particles and the volume, respectively. The Fermi momentum is defined as $k_{\mathrm{F}}=(6\pi^2 n)^{1/3}$.  The Hamiltonian of the system is described by
\begin{equation}
  \mathcal{H}=-\frac{\hbar^{2}}{2m_\mathrm{I}}\nabla_\mathrm{I}^{2}+\sum_{j=1}^{N}\left[-\frac{\hbar^{2}}{2m}\nabla_{j}^{2}+V_{\mathrm{I}j}(r)\right],
    \label{eq:Hamiltonian}
\end{equation}
where the ion induces an atom-ion polarized potential~\cite{Zbigniew15} depending on the distance $r=|\vec r_\mathrm{I}-\vec r_j|$ from an atom to the ion,
\begin{equation}
V_{\mathrm{I}j}(r)=-C_{4}\frac{r^{2}-c^{2}}{r^{2}+c^{2}}\frac{1}{\left(r^{2}+b^{2}\right)^{2}}.
\label{eq:vai}
\end{equation}
The parameters $b$ and $c$ are written in units of $r_{\star}=\sqrt{\left(2m_r C_4 /\hbar^2\right)}$ and $m_r^{-1}=m_{\mathrm{I}}^{-1}+m^{-1}$ the reduced mass and they lock the potential depth and the effective repulsive short-range contribution, respectively. In addition, this pair is chosen to match the $s$-wave scattering length when solving the low-energy two-body problem. The relevant length and energy scales are defined by the competition between the kinetic and the potential interaction range $r_{\star}$. The characteristic energy scale $E^{\star}=\hbar^{2}/2m_r r_{\star}^{2}$ is proportional to the height of the centrifugal barrier for any partial wave and it fixes the upper bound for the energy for isotropic $s$-wave partial wave. For alkali atoms, the range of the potential is very small with respect to the interparticle distance, namely $r_{\star} \ll  k_{\mathrm{F}}^{-1} \sim n^{-1/3}$.  Because of this separation of length scales, a simple yet powerful variational wave function known as the "Chevy ansatz" has been employed successfully to describe Fermi polarons with short-range interactions~\cite{Chevy2006}.

Ion-atom hybrid systems differ significantly from Fermi polarons with short-ranged interactions due to the absence of a hierarchy of length scales, introducing new polaronic effects compared to the neutral case. For typical atom-ion mixtures, the range of the potential~\cite{Tomza2019Review} can be comparable to the interparticle distance, making finite-range effects prominent and leading to a complex interplay between few-body and many-body physics. Such range effects have also recently been shown to be relevant in the neutral case~\cite{pes21}.

Additionally, the strong atom-ion potential can substantially alter the density of the host bath around the ion~\cite{ast21}, thereby modifying the polaronic properties. This density redistribution, as illustrated in \fig{fig1}, is extracted from the ion-atom pair correlation function, which is significantly influenced by the presence of the ion, leading to non-trivial backaction effects.

Our work addresses the many-body problem in \eq{Hamiltonian} using zero-temperature quantum Monte Carlo techniques. In this context, the density distribution of fermions around an ion is governed purely by quantum mechanical effects. The interplay between the attractive ion-atom interaction and repulsive forces—such as short-range repulsion from the potential tail and the Pauli exclusion principle—dominates the behavior of the fermions. The resulting density distribution is shaped by a delicate backaction effect that reduces the fermion density at short distances.

\begin{figure}[htp]
\centering
\includegraphics[width=\columnwidth]{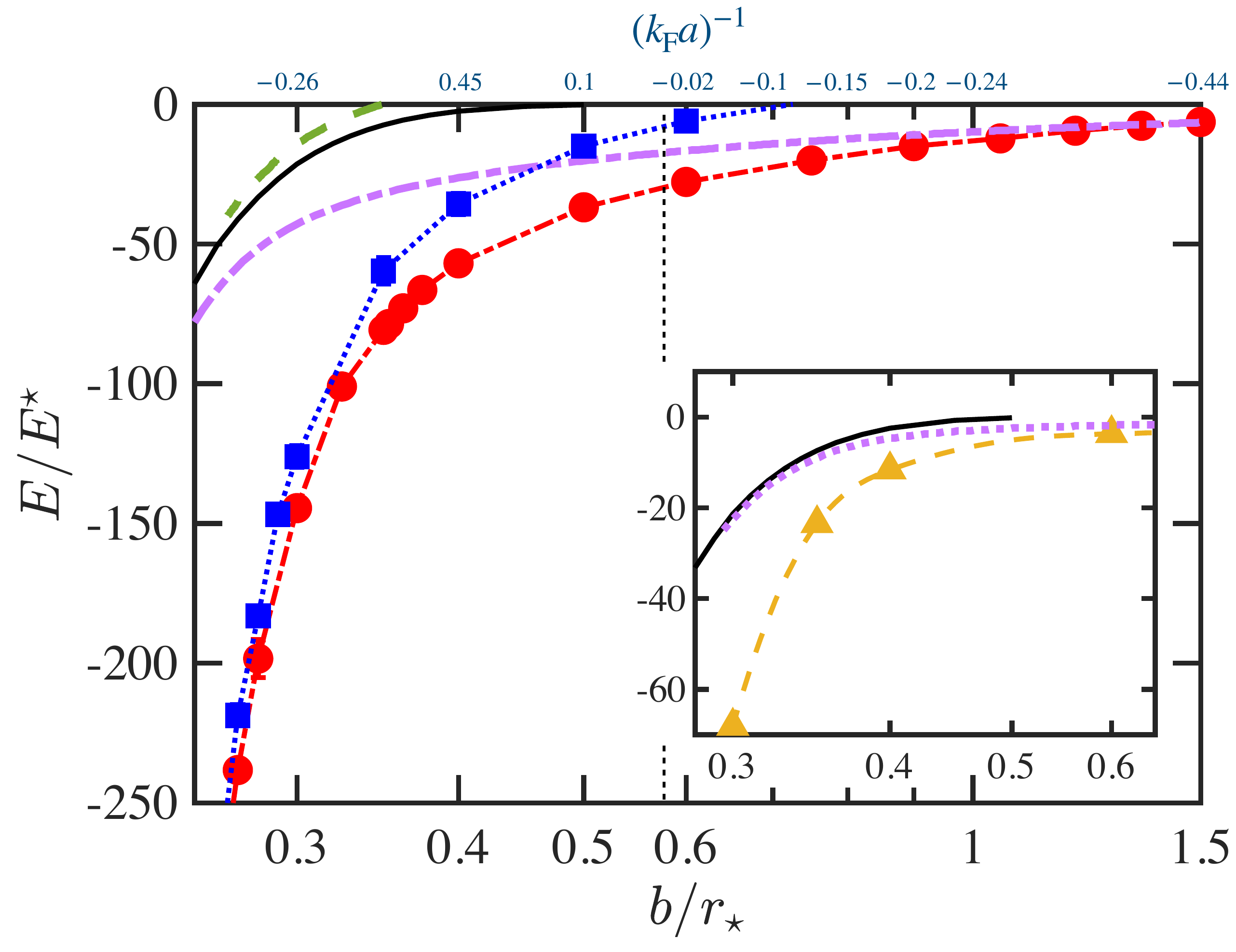}
\caption{Polaron energy $E({\mathbf{p=0}})$ (red circles) and molecular energy $E_{\mathcal{M}}$ (blue squares) as a function of $b$ 
 (also as a function of $\left(k_{\mathrm{F}}a\right)^{-1}$; upper scale) for the unperturbed density \(n r_{\star}^3=1\). Theoretical predictions using the ladder approximation~\cite{chr22} are shown as a purple dash-dotted line for the polaron and a green dashed line for the molecule. Inset: Polaron energies (yellow triangles) at  \(n r_{\star}^3=0.1\). Ladder approximation polaron energy represented by a dotted purple line. The black continuous line depicts the solution for the atom-ion vacuum dimer. Results for $m_\textrm{I}=m$.}
\label{fig:fig2}
\end{figure}

\textit{Quasiparticle properties --} We compute the many-body ground state of \eq{Hamiltonian} as a function of the parameter $b$  for a fixed $c=0.0023r_{\star}$ (see Supplemental Material). For slow impurities, $\mathbf{p}\ll r_{\star}^{-1}$, the polaron energy is computed as $E(\mathbf{p})=\mu+\mathbf{p}^{2}/2m^{\star}$ with $\mu=E(N,1^{k=0}_\mathrm{I})- E(N,0)$, where $E(N,1^{k=0}_\mathrm{I})$ and $E(N,0)$ are the ground-state energies of the Fermi bath with, one ionic impurity in the zero momentum state and without the ion, respectively, while $m^{\star}$ is the effective mass. At the two-body level, we solve the low-energy two-body problem for the potential \eq{vai} in the $s$-wave regime. The wavefunction's asymptotic behavior at large distances provides the $s-$wave scattering length. In particular, for fixed $c\ll r_{\star}$ as in our case, the scattering length has an analytical form,~\cite{szm95,sro20} $a=\sqrt{b^{2}+1}\cot\left[\frac{\pi}{2}\frac{\sqrt{b^{2}+1}}{b}\right]$ ($a$ and $b$ in units of $r_{\star}$). Thus, the two-body spectrum showcases several resonances as a function of $b$. The solution of the two-body problem enters in the explicit choice of the trial wave functions of the Jastrow-Slater form~\cite{SM}.

In Fig.~\ref{fig:fig2}, we plotted the polaron energy as a function of $b$. For values, $b\gtrsim0.58r_{\star}$ the two-body problem gives a scattering solution (no bound states), whereas for  $0.26r_{\star}\lesssim b\lesssim0.58r_{\star}$ the system admits a single two-body bound state.  The energy of the vacuum two-body problem is represented by the black lines in Fig.~\ref{fig:fig2}. Instead, further resonances appear involving high-order few-body states for values $b\lesssim0.26r_{\star}$. We focus on the regime leading up to forming one single-bound state to gain insight into the fundamental polaron-molecule transition. In the scattering regime, $b\gtrsim0.58r_{\star}$ the polaron energy decreases as the resonance is approaching $\left(k_{\mathrm{F}}a\right)^{-1}\rightarrow0^{-}.$  Indeed, as the parameter $b$ decreases, the potential depth increases, resulting in a more tightly bound polaron. This qualitative trend is followed by the ladder approximation approach as well~\cite{chr22}, represented by the purple dashed-dotted line in Fig.~\ref{fig:fig2}. In addition, in the weakly interacting regime $b>r_\star$, both theories agree quite well; however, strong quantitative deviations are observed as $b$ decreases even further. The main reason for the discrepancy is associated with the high nonhomogeneous density bath in the neighborhood of the ion that increases as $b$ decreases. Note that neither backaction effects nor density deformations are fully considered in the ladder approximation. In Fig.~\ref{fig:fig1}(c), we compute the density around the ion and we observe large deviations from the unperturbed density $n$ for length scales on the order of the potential range $r<r_\star$ for different values of $b$.  Yet another intriguing scenario is at the low-density regime, namely $nr_{\star}^3\ll 1$.  Because of the large separations of length scales, one expects to recover the physics of the neutral Fermi polaron at least for $b\gg r_{\star}$.  For both $nr_{\star}^3=0.1$ and $b\gg r_{\star}$ our simulations recover both the ladder approximation~\cite{chr22} and the mean-field (MF) prediction for the energy $E^{\mathrm{MF}}=-\pi^{2}\frac{n}{b}E^{\star}$ ~\cite{note1}.  Interestingly, even for the low densities and the regime where the atom-ion interaction is relevant, $b\lesssim r_{\star}$ deviations are displayed between QMC and the ladder approximation and become more evident as the resonance is approached.  This comparison is shown in the inset of Fig.~\ref{fig:fig2}, revealing that even though the unperturbed density is small, $nr_{\star}^3\ll1$, significant nonperturbative deviations are observed in the region $r<r_{\star}$; see, for instance,   Fig.~\ref{fig:fig1}(d).
\begin{figure*}[htp]
\includegraphics[width=16cm]{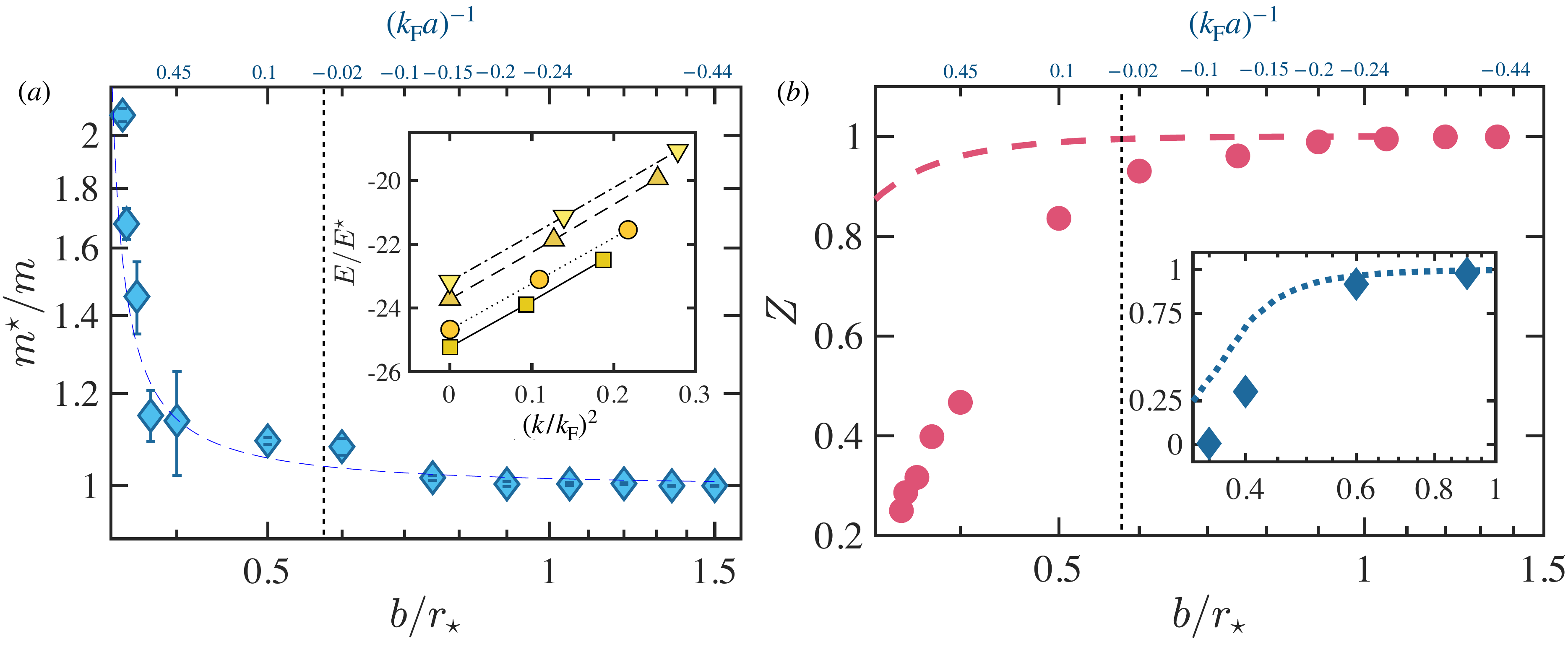}
\caption{(a) Effective mass of the ionic Fermi polaron  for $n r^{3}_\star = 1$. The inset presents the polaron energies as a function of the square momentum vector $\left( k/k_\mathrm{F}\right)^2$ for different numbers of particles in the bath $N = 147$ (yellow squares), $123$ (yellow circles), $93$ (yellow triangles) and $81$ (yellow upside-down triangles). (b) Quasiparticle residue for the density $nr_{\star}^{3} = 1.0$. The dashed and dotted lines are results obtained by Christensen \textit{et al.}~\cite{chr22}. The inset shows results at the density $nr_{\star}^{3} = 0.1$. The vertical dashed lines indicate $b=0.58r_\star$ where $1/a = 0$. Results for $m_\textrm{I}=m$.}
\label{fig:fig3}
\end{figure*}

 In Fig.~\ref{fig:fig2} an excited state known as the "molecular branch" is plotted. The nodal surface is built such that the ionic impurity is injected into the first momentum state above $k_\mathrm{F}$  and exciting one fermion to the corresponding state with opposite impurity momentum, $\mathbf{k}$, such that the total ion-fermion momentum equals zero. The molecular energies are computed with QMC and are represented by the blue squares in \fig{fig2}. The energy is computed by $E_{\mathcal{M}}=E(N-1,1_{\uparrow}^{\mathbf{\mathit{k}>\mathit{k_{\mathrm{F}}}}},1_\mathrm{I}^{\mathbf{\mathit{\left|-k\right|}>\mathit{k_{\mathrm{F}}}}})-E(N-1,0)$,  the difference of the ground-state energies of the system with a molecule formed by the pair $(1_{\uparrow};1_\mathrm{I})$ and the Fermi sea with $N-1$ atoms~\cite{SM}. For a critical value $b_{c}\simeq0.37r_{\star}$ or $\left(k_\mathrm{F}a\right)^{-1}\sim1.28$, the molecular branch meets the polaron state and becomes degenerate to the polaron branch. Akin to the polaron-molecule transition exhibited in the neutral case, the transition is pinpointed around $\left(k_\mathrm{F}a\right)^{-1}\sim1.28$~\cite{Punk2009}; however, for the values explored, our transition does not show any crossing, in contrast to the neutral, zero-temperature and single impurity cases, but instead the polaron-molecule transition is continuous presumably due to the long-range character of the atom-ion interactions and the choice of nodal surface. In the neutral case, it is argued that the first-order phase transition observed in the polaron-molecule is attributed to many-body correlations, whereas the transition vanishes in the few-body limit~\cite{ParishVarenna2023}.  For $b<0.58r_{\star}$, the ion experiences an effective potential consisting of a deep-ion potential supporting a two-body bound state and an additional potential arising from bath deformation.  In other words, introducing an additional fermion attempting to occupy the two-body molecular state results in an energy increase due to the excess of the kinetic energy of the fermions and the  Pauli exclusion; thus the effective potential becomes shallower than the bare potential, preventing the formation of higher-order bound states. Note that this effect arises due to highly local inhomogeneities of the impurity's density -- for example, in the strongly attractive coupled Bose polaron~\cite{SchmidtEnss2022} where the repulsion between bosons is played in our system by the Fermi Pauli blocking. In fact, in Fig.~\ref{fig:fig1} (b) we found that the "molaron" is dressed by roughly $N_{\uparrow} \approx 8$ fermions in the vicinity of the ion.  The molecular branch energy presents notable differences against analytical approaches. Although the ladder approximation relies on a well-educated molecular ansatz, it only provides an upper bound for the molecule energy (see dashed green line in Fig.~\ref{fig:fig2}). In addition, QMC  encompasses all possible correlations in the system, while the molecular ansatz for the ladder approximations is limited to low-order molecule-particle excitation. 

 In Fig.~\ref{fig:fig3}(a), we plot the effective mass as a function of the parameter $b$  for the fixed unperturbed density $nr_{\star}^3=1$.  In the weakly interacting regime $b\gg r_{\star}$, the effective mass is slightly larger than $1$ and increases as one approaches $1/a=0$. On the other hand,  $m^{\star}$ is continuous when entering the two-body sector $b\approx 0.58r_{\star}$. Both effects are consistent with the trend displayed by the polaron energy. As the system becomes more correlated, the effective mass grows rapidly for values  $b<0.58r_{\star}$.  For $b_c$, the effective mass is roughly $m^{\star}\sim 2m$, i.e., the mass of the bare ion-atom molecular pair.  Indeed we find a finite effective mass in this regime consistent with the molaron picture and the number of fermions in the dressed cloud. The rapid increase of the effective mass was also predicted using recent semiclassical approaches~\cite{mysliwy2023longrange}. However, we do not expect any divergence in the range of parameters investigated. From a methodological standpoint, the polaron effective mass \(m^\star/m\) is determined by calculating the dispersion energy of the polaron, $E(\mathbf{p})$, where $k$ is the ion's impurity wave vector, and then extrapolating for a larger number of particles in the thermodynamic limit~\cite{SM}. The inset of \fig{fig3}(a) shows the fits for $b=0.6r^\star$ and $N= 81$, $93$, $123$, and $147$ atoms; in fact, the quadratic behavior of the energies at low momentum demonstrates the reliability of the quasiparticle model.
 
Finally, we plotted in Fig.~\ref{fig:fig3}(b) the quasiparticle weight, which quantifies the difference between the polaron wave function with respect to the initial noninteracting state. The residue is slightly smaller than one for $b>r_{\star}$, namely the ion polaron resembles the noninteracting state.  This result agrees with the ladder approximation by Christensen \textit{et al.}~\cite{chr22} for high and low densities, $nr_{\star}^3=1$ and $nr_{\star}^3=0.1$, respectively.  Nonetheless, for strong coupling and regardless of the density, the residue decays rapidly, approaching zero as one gets close to $b_c$. This behavior is expected because the impurity gets more attracted to the atoms in the bath, and for a critical $b_c$, the molecule becomes the new ground state.   For fixed $b$, a counterintuitive behavior was found first in~\cite{chr22} and also corroborated by our numerical simulations in which the quasiparticle residue increases as the density increases. In our calculations for $b=0.35r_{\star}$ the residue is $Z\approx0.25$ at the density $nr_{\star}^3=1$, while for a smaller density, e.g., $n r_{\star}^3=0.1$, the residue drops to zero within statistical error. In fact, depending on the unperturbed density, the atom-ion potential can trap an average number of fermions as depicted in Fig.~\ref{fig:fig1} (b). Indeed, for a fixed $b$ the number of atoms trapped at high densities is larger than at lower densities. In the latter scenario, the ion interacts with only a few fermions, resulting in relatively significant few-body effects rather than many-body;as a consequence, naively, one expects that the polaron and the dressed dimer become similar and the residue is smaller. Conversely, a sufficient number of fermions are present at higher densities, leading to collective local excitations that dress the ion, building up a tightly bound polaronic state characterized by a higher quasiparticle weight.
        
\textit{Summary and outlook} -- The ionic Fermi polaron quasiparticle is fully characterized by computing the polaron properties. In contrast to the neutral case, the long-range nature of the atom-ion potential can trap a significant number of fermions attracted to the ion, leading to substantial deviations from the unperturbed density and appreciably altering the properties of the polaron. In fact, our results quantitatively differ from the analytical approaches such as ladder approximation, for which the density is taken constant and the backaction effects are ignored. We also unveil a smooth polaron-molecule transition at zero temperature, starkly contrasting with the neutral case. The quasiparticle residue can be accessed by Raman spectroscopy with momentum-dependent measurements~\cite{Ness2020}, while the effective mass can be extracted by diffusive transport of the ions as a function of an external electric field ~\cite{Dieterle2020b} or alternatively by measuring the low-lying compression modes~\cite{nas09} in the high imbalance atom-ion mixture trapped in a controllable Pauli trap. An interesting avenue to pursue in the future is to study excitonic polarons, which arise from the interaction of excitons and a Fermi gas in 2D~\cite{Campbell2022,Efimkin2021}.\\

\textit{Acknowledgements --}  We thank G. Bruun, S. Giorgini, A. Camacho, K. Myśliwy, K. Jachymski and S. Pilati for insightful discussions. We thank E. Christensen for sharing his results in the ladder approximation. S.A.V acknowledges financial support from the Brazilian agency,
Funda\c{c}\~ao de Amparo \`a Pesquisa do Estado de S\~ao Paulo
grant \#2023/07225-0, São Paulo Research Foundation (FAPESP).
Computational resources have been provided in part by CENAPAD-SP at Unicamp.
R.P thanks the computer support from LaMCAD/UFG and L.A.P.A acknowledges financial support from PNRR MUR project PE0000023-NQSTI.

\bibliographystyle{apsrev4-2}

\begin{thebibliography}{86}
\bibitem [{not()}]{note1}%
  \BibitemOpen
  \href@noop {} {}\bibinfo {note} {The mean-field energy corresponds to the
  limit $b\gg r_{\star}$, however it can be also recovered from first-order
  perturbation theory $E^{\mathrm{MF}}=nV(\mathbf{k=0})$ being $V(\mathbf{k})$
  the Fourier transform of the ion-atom potential.}\BibitemShut {Stop}%
\end{thebibliography}

\renewcommand{\theequation}{S\arabic{equation}}
\renewcommand{\thefigure}{S\arabic{figure}}
\renewcommand{\thetable}{S\arabic{table}}

\onecolumngrid

\newpage

\setcounter{equation}{0}
\setcounter{figure}{0}
\setcounter{table}{0}


\section*{Supplemental Material for  "Fermi polaron in an atom-ion hybrid  systems" }
\begin{center}
Renato Pessoa,   Silvio Vitiello  and Luis A.~Pe{\~n}a~Ardila, \\
\end{center}

This supplementary material describes details on the two-body problem for the atom-ion interaction potential, employing the Buckingham polarization potential model where analytical expression can be derived for the $s-$wave scattering length. In addition, we include the most relevant systematic and technical aspects of the variational  (VMC) and diffusion quantum Monte Carlo  (DMC) methods. Specifically, we discuss the detailed functional forms of our trial wave functions and the estimators employed for the rigorous characterization of ground-state properties.\\

\maketitle

\onecolumngrid

\section{I. The atom-ion interaction}
\label{app:ai}

The interaction between a neutral atom  and a positive charged particle is given by~[16]

\begin{equation}
V_{\mathrm{ai}}(\mathbf{r})=-\frac{C_4}{r^4}.
\label{eq:vai}
\end{equation}
The strength of this potential
$C_4=\alpha e^2 / 2$ depends on the static polarisability
$\alpha$ of the atom and $e$ is the elementary electrostatic charge. The electrostatic forces, even playing an important role in the behavior of the system, are not sufficient to fully describe
the atom-ion interaction.  At short ranges, the exact form of the potential is complex and generally unknown, and the $V_{\rm ai}$ potential is singular as $r \rightarrow 0$.
A common approach~[67,70] to address these difficulties is provided by a convenient regularization of this potential~[73],
\begin{align}
    \label{eq:kry}
V_{\mathrm{ai}}^r(\mathbf{r})
    =-C_4 \frac{r^2-c^2}{r^2+c^2} \frac{1}{\left(b^2+r^2\right)^2},
\end{align}where $b$ and $c$ are parameters chosen to describe the low energy
properties of the system. The cut-off radius $b$ deepens the modeled
potential as its value is decreased, while $c$ sets the distance at which the interaction becomes repulsive, to
avoid overlap of the electronic wavefunctions of the atom-ion dimer. The pair ($b,c$)
parameters draw the  two-body energy spectrum of the system parametrized by the atom-ion
scattering length. Additionally, the regularized potential energy has a
finite value as $r\rightarrow 0$, a useful feature in many numerical
calculations.\\

For the alkali atoms, the characteristic length is $r_{\star} \sim {\cal O}(10^3)$~\AA, and in this work, the value $c=0.0023r_{\star}$ is used. This choice mimics the experiment with $^{138}\textrm{Ba}^+$ ions immersed in a bath of $^6\textrm{Li}$ atoms~[63] and is used in the analytical calculation by Christensen \textit{et al.}~[70]. Changing the
parameter $b$ tunes the energy of the bound state, mimicking
experiments where Feshbach resonances have recently been observed~[63].  From a two-body level, we determine the values of $b$ at which the dissociation of bound states of the atom-ion occurs by solving the Schrödinger equation for the zero-energy limit. Namely,
\begin{align}
\left[
    \frac{d^2}{d r^2}+\frac{r^2-c^2}{r^2+c^2} \frac{1}{\left(b^2+r^2\right)^2}
    \right] \psi(r)=0.
    \label{eq:sf0}
\end{align}
To obtain the values of $b$ for dissociation, the wave function must satisfy the boundary conditions $\psi(0)=0$ and $\psi^{\prime}(0)=\epsilon$ at the origin, where $\epsilon$ is any small number. The $s$-wave scattering length is determined by iteratively solving the radial time-independent Schrödinger equation of \eq{sf0} for different values of $b$. The solution for the radial function, $\chi(r) = \psi(r)/r$, can be found by matching the semiclassical wave function in the attractive part of the interaction with the asymptotic exact solution using the $-1/r^4$ potential~[59]. It is possible to assume that as $r\rightarrow \infty$, the potential approaches zero, leading to
\begin{align}
    \chi(r)  \underset{r\rightarrow \infty}{\sim}  A(r-a),
\end{align} where $A$ is a constant and $a$ is the scattering length for the $s$-wave zero-energy. The scattering length exhibits resonances as $b$ decreases, each one corresponding to the appearance of a deeper bound state.

\begin{figure}[ht]
\includegraphics[width=0.8\columnwidth]{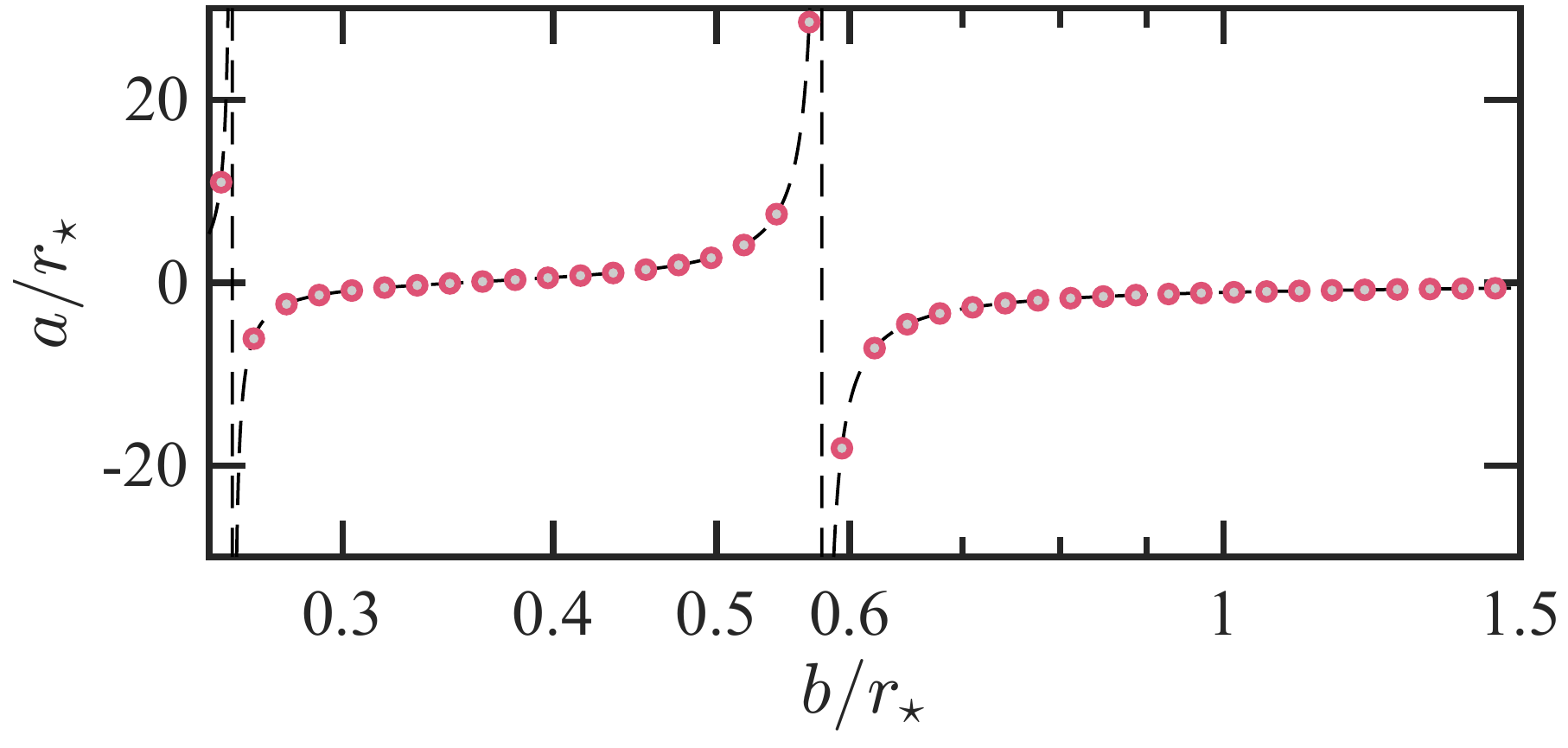}
    \caption{Numerical solution of the atom-ion scattering length as a function of the parameter $b$  and $c=0.0023 r^{_\star}$ (red circles) and exact result Eq.~\ref{eq:aofb}.}
    \label{fig:Fig1S}
\end{figure}

We can confirm that the constant prefactor dependent on $c$ in the regularized potential described by Eq. \eqref{eq:kry}, responsible for its short-range behavior, can be safely ignored in this work by considering the Buckingham polarization potential. This potential is obtained when the approximation
    $(r^2-c^2)/(r^2+c^2)\approx 1$
is valid.  In this case, the Schrödinger equation can be solved analytically and the scattering length can be expressed as~[76,77,80] 
    \begin{align}
        \label{eq:aofb}
        a = (1+b^2)^{1/2}\cot\frac{\pi}{2}(1+b^{-2})^{1/2}.
    \end{align}
In \fig{Fig1S}, we superimpose the scattering lengths of Eq. \eqref{eq:aofb} with those obtained by iteratively solving the zero-energy Schrödinger equation with the full regularized potential of Eq. \eqref{eq:kry}. The agreement between the two is excellent.

\section{II. Quantum Monte Carlo Details}
\label{app:QMC}

 In our simulations, the antisymmetrical nature of wave functions describing Fermi particles is accounted for by imposing the fixed-node approximation used in the diffusion Monte Carlo method. This method considers a given a priori nodal structure for the system and gives the best approximation at zero temperature for the ground state energy consistent with the chosen nodal structure.  

The wave function model used in the simulations is of Jastrow-Slater form,
\begin{align}
\psi_T(R) = \prod_{i} f(r_{i\mathrm{I}})D_{\uparrow}
e^{i \vct k \cdot \vct r_\mathrm{I}},
\label{eq:JS}
\end{align}
where $R\equiv\{\vct r_1,\ldots \vct r_N, \vct r_\mathrm{I}\}$ represents the space points of all the atoms, and $D_{\uparrow}$
is a Slater determinant for the up-spin particles. The single particle orbitals are chosen to be plane
waves. Wave vectors are given by $k = \frac{2\pi}{L}\sqrt{n_x^2 + n_y^2 +
n_z^2}$ where $n_x$, $n_y$ and $n_z$ are positive or negative
integers and $L$ is the side of the simulation box. In this way, 
\begin{equation}
    D_{\uparrow}=\mathcal{A} \prod_{j} e^{i \vct k \cdot \vct r_{j}}
\end{equation}
where $\mathcal{A}$ is an antisymmetrizer operator. Most calculations considers closed shells for the up-spin bath, where the number of particles is defined by the maximum value of $I=n_x^2 + n_y^2 +
n_z^2$. For example, for the bath with $N=81$ particles, we have $I_{\textrm{max}}=6$ to take into account the closed shells case.\\

The form of the Jastrow
term $f(r)$, in the trial wave function, depends on the value of $b$. For $b/r{_\star} > 0.58$ the
lowest-order constrained variational (LOCV)
method~[81-83] is employed to obtain $f(r)$. This involves solving the two-body
Schr\"odinger like equation
\begin{equation}
\label{eq:locv}
    \left[ -\frac{\hbar^2}{2m_r}\nabla^2 + V(r)\right] f(r) = \lambda f(r),
\end{equation}
where the $\lambda$ parameter is chosen to ensure the
continuity of $f(r)$. Specifically, $\frac{df}{dr} = 0$ for $r=\mathcal{D}$, and it is imposed that $f(r\geq \mathcal{D})=1$; $\mathcal D$ is the range of the Jastrow term. The value of the $\mathcal D$ parameter is obtained by minimizing the value of the energy in the VMC as explained further below.

For $b/r{_\star} < 0.58$, the interatomic potential becomes deeper and it is important to consider a phononic contribution in the wave function. The Jastrow factor obtained from \eq{locv} is modified by adding a tail
\begin{align}
    A\exp \left [ -C \left( \frac{1}{L-r} +\frac{1}{r} \right )\right ]
\end{align}
for $\bar{r} \le r \le L/2 $, where $\bar r$ is a variational parameter chosen to minimize the trial energy.  Parameters $A$ and $C$ are chosen to ensure continuity of the wave function at the distance $\bar{r}$ and in the simulation box boundary. 
Simulations considering the phononic contribution have shown that for large enough values of $\mathcal D$ its value is not critical, we have adopted ${\mathcal D}=L/2$.

The VMC method involves sampling the probability distribution
\begin{equation}
p(R) = 
\frac 
{ |\psi_T (R)|^2 }
{ \int dR \ |\psi_T (R)|^2 },
\end{equation}
where $\psi_T(R)$ is the trial wave function. The variational energy is estimated by approximating the integral $\int dR\, p(R) E_L(R)$ by the average values of the local energy $E_L(R)=H\psi_T/\psi_T(R)$ over the sampled configurations~[84]. The aim of VMC in this work is to optimize the variational parameters entering the trial wave functions. 

The DMC method solves the Schrödinger equation for imaginary time $\tau$
\begin{align}
    -\frac{\partial \Psi}{\partial \tau} = (H-E_T) \Psi,
\end{align}
where $E_T$ is a constant. Importance sampling is required for the system to explore the most important regions of the configuration space. Through this transformation, converged samples are drawn from the distribution $\psi_0\psi_G$, where $\psi_G$ is the trial wave function $\psi_T$. The Hermiticity of the system Hamiltonian allows for an unbiased estimation of energy in systems following Bose-Einstein statistics by averaging the local energy. However, estimation of any quantity $Q$ that does not commute with the Hamiltonian requires extrapolation $Q_{\rm extr}=2Q_{\rm DMC} - Q_{\rm VMC}$, which depends on a variational $Q_{\rm VMC}$ result.\\

Simulations are carried out by considering the fixed-node approximation to avoid the sign problem, a change in the sign problem when one particle crosses the nodal surface of the guiding function~[85,86]. In the simulations, to guarantee results free from time step bias, the value of the time step was carefully analyzed by verifying that the energy of an impurity in a bath of $N=123$ particles did not change for $\Delta \tau = 10^{-4}$, $\Delta \tau = 2\times 10^{-4}$ and $\Delta \tau = 5 \times 10^{-4}$. The value $\Delta \tau = 10^{-4}$  was chosen to avoid a move where a particle crosses more than one nodal hyper-surface and reaches a region with the original sign, the cross-recross error.\\

\begin{figure}[ht]
    \centering
   \includegraphics[width=0.8\columnwidth]{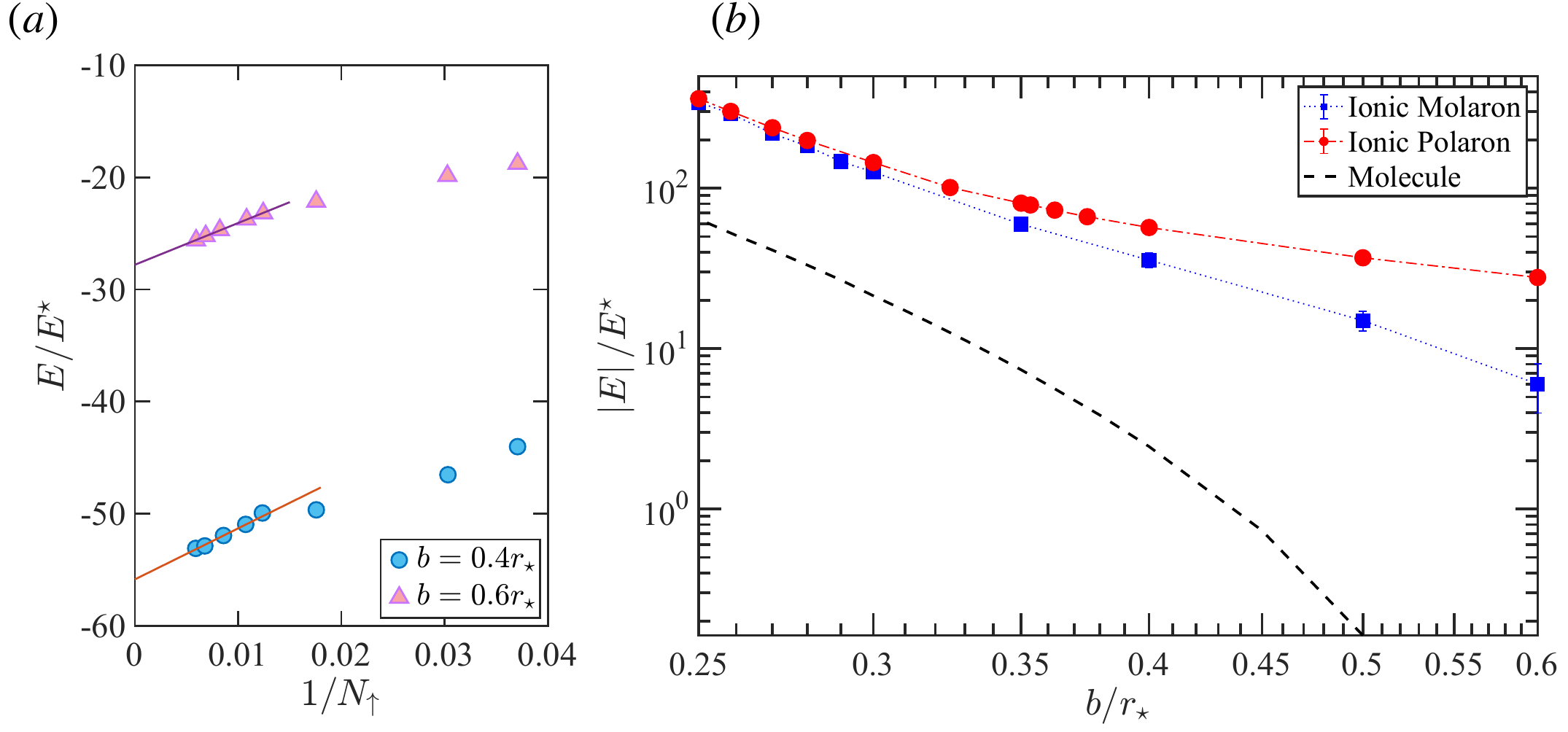}
    \caption{\textbf{(a)} Typical size effect for the  energy at zero momentum for two values of the atom-ion potential. \textbf{(b)} Polaron energy  of the attractive polaron ($E_{p}$) and the molecular polaron "molaron" ($E_{\mathcal{M}}$) in the neighborhood of the resonance. The same parameters are from Fig. 2 in the main text.}
    \label{fig:FigS2}
\end{figure}

The wave function associated with the polaron is constructed following the Jastrow-Slater model described in \eq{JS} where the momentum states of the up-spin particles of the bath are fully occupied for $I\le I_{\textrm{max}}$. On the other hand, the wave function for the molecular case has a different nodal structure that emerges from changes made in the Slater determinant. In this case, the down-spin impurity is set in the first momentum state above the Fermi state. Furthermore, a bath particle is removed from the highest filled momentum state and set in the same ion impurity momentum state. As observed in Fig.2 in the main text, in the neighborhood of the resonance of the two nodal surfaces match despite the molecular nodal surface imposed in the trial wave function should favor the molecule over the polaron. In Fig.~\ref{fig:FigS2} we have extended further the parameter window of $b$ to observe that the polaron and molecular branch are essentially indistinguishable within statistical error. \\

Another possibility is to use a BCS trial wavefunction that might lower the energy. However, this change is very small for the N+1 problem considered in our system as it correlates only one pair orbital, while other correlations with the bath are unpaired, akin to the normal phase of the gas. In addition, the appropriate choice of the Jastrow factor is essential to provide the correct correlation between the ion impurity and the bath particles. The solution two-body Schrödinger equation indicates the bound state for $b < 0.58 r_{\star}$, as shown in Fig. 2 in the mean text. This bound state physics is also inspired by the ansatzs in~[87] and subsequently exploited also for the atom-iom hybrid system in~[68].

The polaron energy is estimated by subtracting from a simulation with one ion in the momentum state with $\mathrm{k}=0$ and the pure system formed by the bath (as defined in the main text)
\begin{align}
\mu=E^0(N,1_\mathrm{I})- E(N,0).
    \label{eq:epol}
\end{align}

Although $E(N,0)$ is known exactly, the energy value for the pure system was obtained through a simulation with the attempt to reduce size effects. However, $\mu$ is not an easy quantity to estimate, the long-range character of the interatomic interaction requires large systems, \textsl{i.e.}, a large number of particles in the bath, that tend to mask the contribution of the impurity in the uncertainty of the simulation result. Nevertheless, a careful consideration of size effects in the results was performed, see panel (a) in Fig.~\ref{fig:FigS2}.

The effective mass $m^*$ of the ionic Fermi polaron obtained for the model of quasiparticle is given by
\begin{equation}
     E(\mathbf{p})\sim\frac{\mathbf{p}^{2}}{2m^{*}}+\mu, \label{eq:quasiE}
\end{equation}
where $\mathbf{p}=\hbar\mathbf{k}$ is the impurity momentum and $\varepsilon$ the ground state polaron energy. We extracted this quantity by placing the impurity in the first three states available in three different simulations. Finally, by plotting $E_{p}(\mathbf{k})$ as a function of $k^2$, with a fit to the results, the effective mass was obtained. A different method was used in Ref.~[69] where this quantity could be obtained in a single run by considering the imaginary time displacement of the impurity.

At zero temperature, an ideal Fermi gas has all its lowest energy levels filled. The introduction of an impurity induces low-energy excitations, which might result in quasiparticles. The Migdal theorem~[89] asserts the existence of a discontinuity in the momentum distribution $n(k)$ at the Fermi momentum $k_\mathrm{F}$ of the ion, characterizing the residue $Z$. This residue is calculated as the difference between the momentum distribution,
\begin{align}
    Z = n(k_\mathrm{F} - \delta) - n(k_\mathrm{F} + \delta),
\end{align}
where $\delta$ is an infinitesimal positive constant. In the thermodynamic limit,
$n(k_\mathrm{F}+\delta)$ approaches zero due to the scaling of the associated one-body density matrix with the system volume. Consequently, the residue $Z$ can be identified with the momentum distribution at $k_\mathrm{F}$ of the ion. An estimator for $Z$ is obtained from the one-body density matrix
\begin{equation}
    Z = \lim_{\vert\vct{r}'_\mathrm{I}
    -\vct{r}_\mathrm{I}\vert\rightarrow\infty}
    \left \langle \frac{\psi(R')}{\psi(R)}
    \right \rangle
    \label{eq:z}
\end{equation}
where $R'\equiv\{\vct r_1,\ldots \vct r_N, \vct r'_\mathrm{I}\}$ represents space points for atoms in the bath, and $\vct{r}'_\mathrm{I}$ represents an arbitrary displacement of the impurity. A non-zero residue characterizes the polaron as a quasiparticle, indicating the overlap between the ground state wave function of the non-interacting (impurity and free gas) and the interacting system~[88,90].

\subsection{Correlation function and density}

\begin{figure}[ht]
    \centering
    \includegraphics[width=1.0\columnwidth]{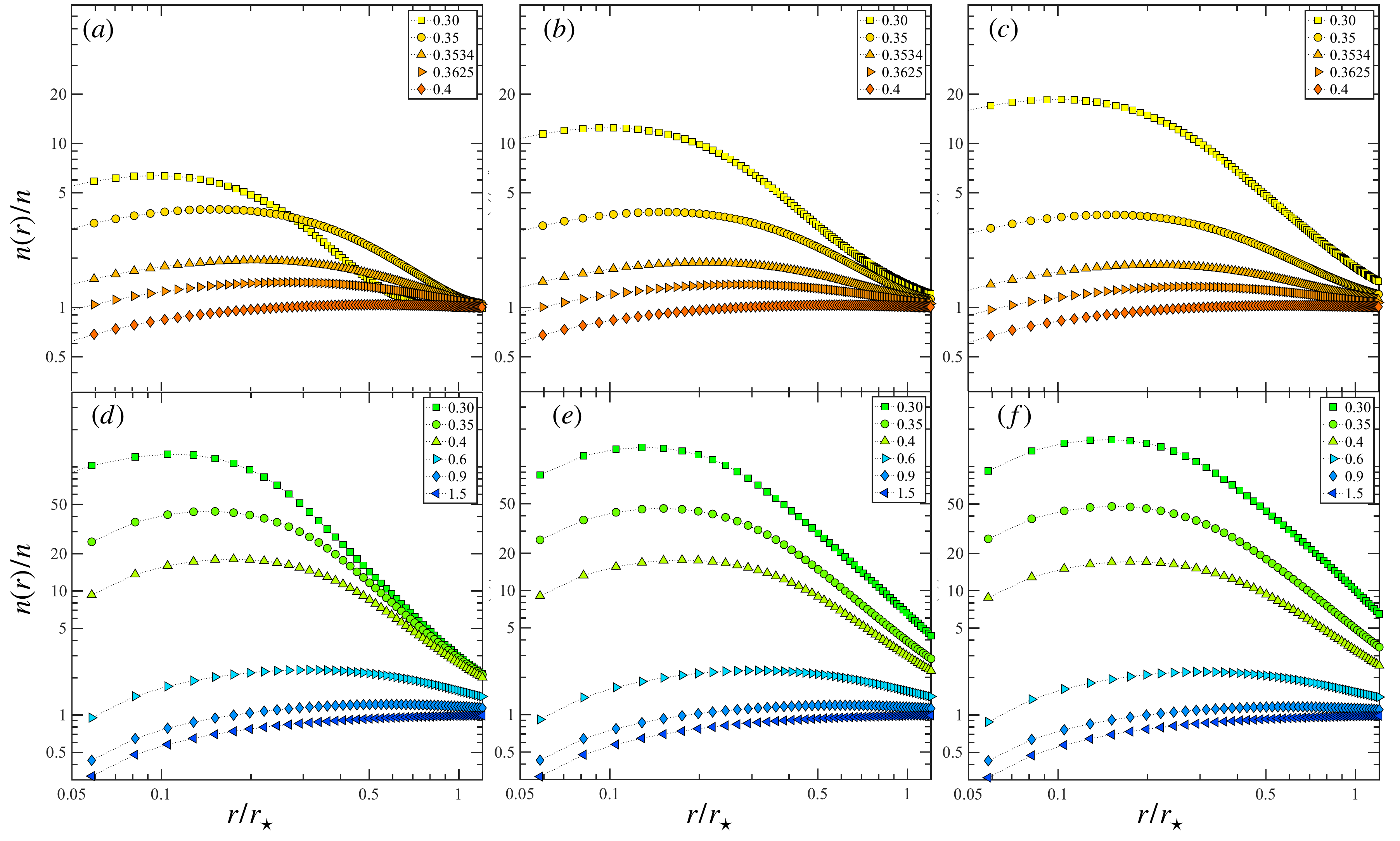}
    \caption{(a-c) Density $n(r)$  computed with $\mathrm{VMC}$, $\mathrm{DMC}$ and extrapolated (see Fig.1 in mean text) respectively for $nr_{\star}^3=1$. (d-f) Same calculations for the $\mathrm{VMC}$, $\mathrm{DMC}$ and extrapolated densities for $nr_{\star}^3=0.1$.  For large distances $r \gg r_{\star}$, the density converges to the unperturbed density, $n(r)/n \rightarrow 1$.
}
    \label{fig:FigS4}
\end{figure}

From the ion-fermion pair correlation function $g_{\mathrm{IF}}(r')$, we compute the number of fermions around the impurity at a distance $r$ by computing $N_{F}(r)=4\pi n\int_{0}^{r}dr'r'^{2}g_{\mathrm{IF}}(r')$ and the density around the impurity similarly by $n(r)=n\frac{\int_{0}^{r}dr'r'^{2}g_{\mathrm{IF}}(r')}{r^{3}/3}$~[91].\\

In Fig.~\ref{fig:FigS4} we plotted the density deformation $n(r)$ as a function of the distance for different parameters $b$ using the VMC, DMC and extrapolated estimator (see also Fig.1 mean text). We observe that in general the differences between the VMC, DMC and the extrapolated estimator are small, except in the case of density $nr_{\star}^3=1$ and $b=3.0r_{\star}$. The horizontal range is chosen $r/r_\star \approx 1$ to better illustrate the proper behavior around the ionic impurity. However, for large distances $r \gg r_{\star}$, the density converges to the unperturbed density, $n(r)/n \rightarrow 1$. Backflow correlations in QMC might provide lower energy, however we do not expect a considerable change as as we are still in dilute regime

\section{III. INFINITE MASS CASE}
\label{app:IMC}

\begin{figure}[ht]
    \centering
    \includegraphics[width=0.5\columnwidth]{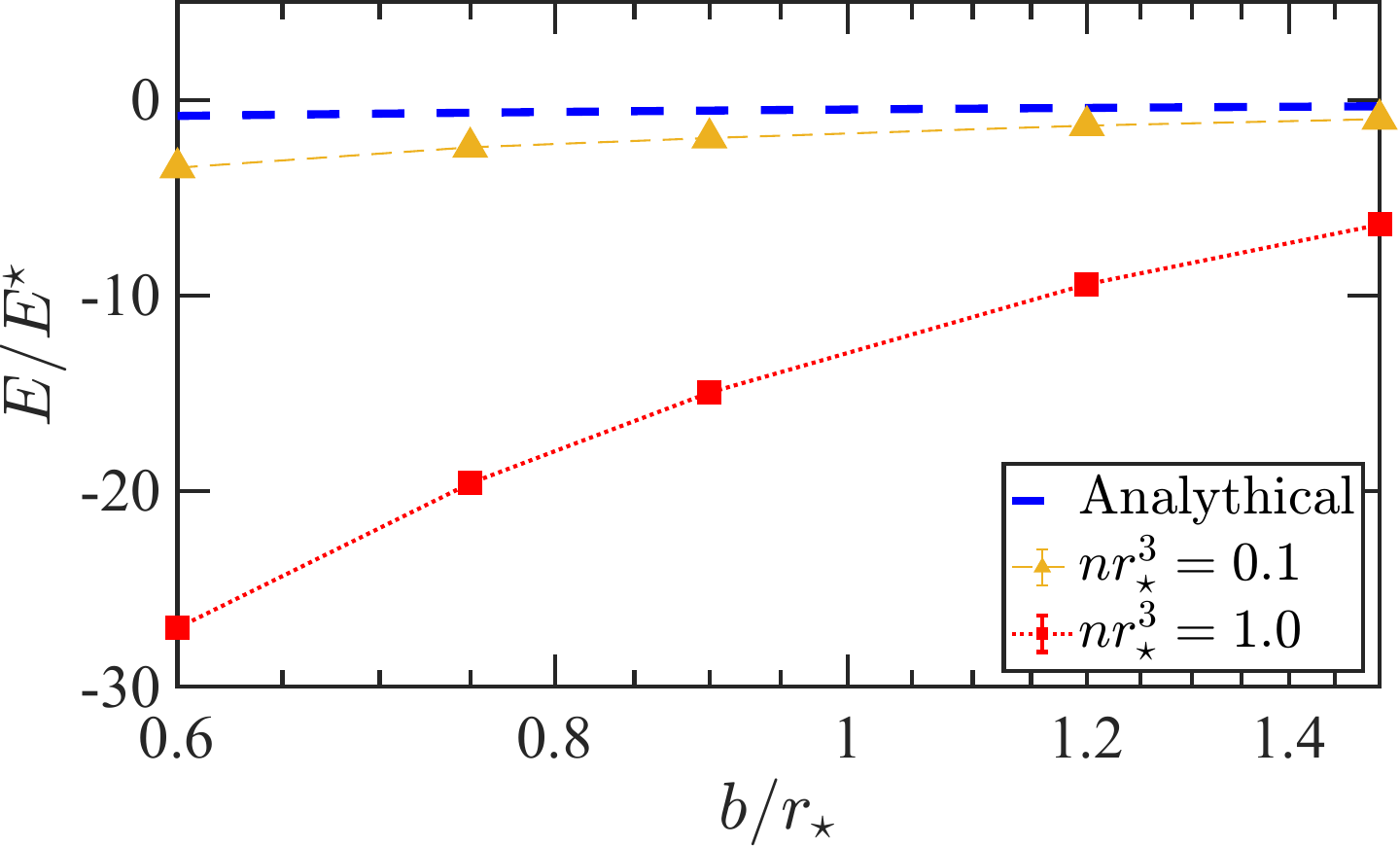}
    \caption{Polaron energy for the relevant experimental case $^{176}\mathrm{Yb^{+}}-{}^{6}\mathrm{Li}$}
    \label{fig:FigS3}
\end{figure}

One possibility to reduce micromotion and characterize the Fermi polarons is to consider the regime of heavy impurities. In Fig.~\ref{fig:FigS3} we estimate the polaron energy and the quasiparticle residue for the relevant experimental case $^{176}\mathrm{Yb^{+}}-^{6}\mathrm{Li}$ which corresponds to the mass ratio, $m_{\mathrm{I}}/m\sim30$. The polaron energy is plotted for two densities, $nr_{\star}^3=0.1$ and  $nr_{\star}^3=1$. In the former case and for $b\gg r_{\star}$ the results are compatible with the mean-field result for the heavy Fermi polaron, $E=-\pi^{2}\frac{n}{b}\frac{\hbar^{2}}{2m_r r_{\star}^{2}}$.

\end{document}